# Smart City Digital Twin Framework for Real-Time Multi-Data Integration and Wide Public Distribution

Lorenzo Adreani [a], Pierfrancesco Bellini [a], Marco Fanfani [a], Paolo Nesi [a, *], Gianni Pantaleo [a]

[a] University of Florence, DISIT Lab, Department of Information Engineering, Via S. Marta 3, Firenze, Italy, 50139

www.disit.org, www.snap4city.org

**\* Corresponding author email address**: paolo.nesi@unifi.it

**Abstract**

Digital Twins, that are as faithful as possible digital replica of real entities, enabled by rapid advances in Big Data and IoT/WoT (Internet/Web of Things) technologies, are becoming fundamental tools to monitor and control the status of entities, predict their future evolutions, and simulate alternative scenarios to understand the impact of possible changes. More recently, Digital Twin solutions have been applied in the context of Smart Cities. Thanks to the large deployment of sensors, together with the increasing amount of information available for municipalities and government organizations, it is possible to build accurate virtual reproductions of urban environments including structural data and real-time information. Such solutions can undoubtfully help city councils and decision makers to face future challenges in urban development and improve the citizen quality of life, by analysing the actual conditions, evaluating in advance through simulations and what-if analysis the outcomes of infrastructural or political changes, or predicting the effects of humans and/or of natural events. In this paper, the Snap4City Smart City Digital Twin framework is presented, which is capable to respond to the requirements that were identified in the literature and by the international forums. Differently from other state-of-the-art solutions, the proposed architecture provides an integrated solution for data gathering, indexing, computing and information distribution offered by the Snap4City IoT platform, therefore realizing a continuously updated Digital Twin of the urban environment at global and local scales. 3D building models, road networks, IoT devices, WoT Entities, point of interests, routes, paths, etc., as well as results from data analytical processes for traffic density reconstruction, pollutant dispersion, predictions of any kind, what-if analysis, etc., are all integrated into a freely accessible interactive web interface, a key aspect to support the citizens participation in the city decision processes. Additionally, our solution can provide tools to address the What-If analysis to let the user performs simulations and observe possible outcomes. As case of study, the Digital Twin of the city of Florence (Italy) is presented. The solution, bundled with the Snap4City platform, is released as open-source, and made available through our GitHub repository (https://github.com/disit) and as docker compose.

## 1. Introduction

The concept of digital twin was first introduced in 2002 by M. Grieves [1], [2] originally named *Mirrored Spaces Model* [3]. During the years the idea was refined and renamed, till J. Vickers of NASA coined the term *Digital Twin* [4]. A Digital Twin can be described as digital replica of a physical entity. First applications of the Digital Twin concept can be found in the aircraft and aerospace industry. More formally, as reported in [5], "*A Digital Twin is an integrated multiphysics, multiscale, probabilistic simulation of an as-built vehicle or system that uses the best available physical models, sensor updates, fleet history, etc., to mirror the life of its corresponding flying twin*". Later, Digital Twins begun to emerge in other fields, such as, for example, in the manufacturing [6, 7], or in construction industries [8]. The increased adoption of Digital Twin was enabled by technological advances and new paradigm such as Internet-of-Things and Web-of-Things (IoT/WoT) [9], Big Data [10], and Industry 4.0 [11]. Indeed, the information gathered from sensors and devices, and the capability to manage and process huge amount of information [12] are key elements in the development of faithful Digital Twins.

More recently, the concept of Digital Twin has started to be applied in the context of cities [13], in particular for Smart Cities environments and tools [14]. Indeed, according to recent studies on urbanization [15, 16, 17], most of the worldwide population lives in urban areas, and projections indicate that urban population will grow in the near future. It is undoubtful that such increasing urbanization will pose several challenges to decision makers to guarantee satisfactory quality of services at city users. Such challenges span in several fields, from mobility and transport organization [18, 19] to environmental and energy management [20, 21], to urban planning [22].

In this context, Smart City Digital Twins (SCDTs) are a fundamental tool required to assess the status of the urban environment and to perform analysis and simulations to guide future strategic development and plans. Initially most of the research focused on the construction of 3D City Models [23], three-dimensional representation of the city that can have different Level of Details (LoD) [24]. Clearly, since such 3D models must cover city-wide areas, both the production of the 3D models, as well as their handling and processing pose a challenging task still to be solved [25]. However, a SCDT must go beyond, not being limited to 3D representations of the urban area at local and global scales, and including real-time data, enabling different kind of operations such as simulations and predictions, and help to increment the involvement of citizens though participatory processes [26].

In [13], the authors identified three main layers composing a SCDT: a first layer included the heterogeneous data types such as buildings, maps, data from sensors, etc., realizing the so-called City Information Model (CIM). The second layer aimed to cover basic functionalities, communications and can implement analytics and software. This can be achieved with an adequate IoT/IoE platform able to manage and manipulate huge amount of data. The final layer was supposed to be devoted to visualization and distribution thought 3D engines and web applications. In order to guide the SCDT definition and development, a series of requirements were identified [27]. Requirements have been divided in three main groups: Data, Interactivity, and Interoperability. The first group responds to the question about which data a SCDT must include. These encompass the 3D representation of the city such as buildings and other urban entities, together with representations for sensors, services, heatmaps, specific areas or paths. Terrain elevation is considered since it provides information for building ground positioning, and orographic aspects of the city and its surroundings. The second group of requirements focus on the interactive functionalities that the SCDT interface must provide. Starting from essential 3D map controls (to change the point of view), they include picking and manipulation functionalities fundamental to provide a valid user experience. Finally, the third group describes requirements on interoperability that should be met to guarantee accessible, integrated, replicable, and affordable SCDT solutions, able to ingest data from different sources and with different formats, and to provide data analytic services. These requirements are aligned with the main technical challenges reported in [26]: data accuracy and availability are fundamental to represent all the required information in a SCDT, also considering the dynamicity and the amount of data gathered; interoperability aspects are considered at levels of data format, software, and systems compatibility; free licences should lead to guarantee the maximum accessibility to the solution and avoid technological lock-in on proprietary solutions.

In this paper, Smart City Digital Twin framework developed on top of the Snap4City Platform is described [28]. Snap4City is an official FIWARE platform, EOSC platform, and a set of libraries on Node-RED [29], [30], and it is at present in operational use on several installations which are also federated each other. The Snap4City platform is able to manage multiple tenants and billions of data with the key focus on interoperability. The research has been performed in the context of MOST national center on Sustainable Mobility in Italy [31]. Snap4City framework is applied in different Smart Cities and areas: Italian (Firenze, Pisa, Livorno, Prato, Lonato del Garda, Modena, etc.) and European cities (Antwerp, Santiago De Compostela, Valencia, PontDuGard-Occitanie, Dubrovnik, Mostar, and West Greece, etc.), and their surrounding geographical areas. The largest installation of the platform is multi-tenant, managing advanced Smart City IoT/IoE applications with 20 organizations, 40 cities and thousands of operators and developers, including free trial and test which increases complexity.

*The main contributions of the paper and its organization are reported at the end of Section 2.*

## 2. Related Works

SCDTs are quite novel, however in the recent period some solutions begun to appear due to the increasing interest in the topic with public or private commitments. Unfortunately, to the best of our knowledge, only a limited number of proposed SCDT solutions released technical documentations, papers or provided freely accessible interfaces to study and assess their solutions. Therefore, in this Section, a selection of the most representative solutions is reported, to give a general context on SCDT development endeavours. Then, for those solutions offering an accessible interface a deeper analysis has been conducted enabling us to assess their compliance with respect to the identified requirements reported in **Tables 1**, which are a wide extension and reorganization of those of [27]. Each requirement is named and described in the table.

The identified requirements have been classified in three groups:

- **Field Interoperability** which refers to the needs devoted to data ingestion/gathering, data transformation, data storage and modelling, IoT network management and interoperability, exploitation of API, federation of API and platforms, etc. This means that this group refers to data ingestion, event driven data, and how to produce data to act on the physical counterpart of the digital twin local and global.
- **Data and computing for representation** group refers to the needs of modelling and integrating complex data, and the related massive computing for higher level data types which are: predictions, traffic flow, heatmaps, trajectories, 3D representations, alarms, origin destination maps, etc. These data kinds are typically not accessible from third party and have to be directly computed on platform to produce data / insight which need to be represented in real time in the front end.
- **Distribution and interaction**. Former data and computed representation need to be prepared and distributed in a suitable manner to be accessible at the front end in which the DT is presented to the decision maker, and on which some interaction must be possible to close the loop and act on the field. For example, to send decisions, changes configuration on the IoT network, what-if analysis to simulate effects and take decisions.

Efforts to produce SCDTs have been undertaken by several cities around the world [13]. For example, Helsinki [32, 33, 34], Gothenburg, Paris, Rennes [35], London, New Castle, Rotterdam [36], Berlin [37], Stockholm [38], Zurich [29], Herrenberg [40], Munich in Europe; Toronto [41], Boston [42] in Nord America; Amaravati, Singapore [43], Shanghai [44] in Asia; Wellington [45] in Oceania.

Most of these solutions seems to be work-in-progress or very limited case of studies and provide a very limited capabilities for Field Interoperability layer, not providing an open access to inspect and test the SCDT, and neither their 3D representation of the city in some interactive manner. If sometimes this is due to privacy policies or licence limitations that do not allow the free distribution of the SCDT, in other cases the lack of access could be related to difficulties to provide solutions which may work with regular clients' solutions (e.g., web browser), instead they would require the usage of dedicated 3D engines requiring specific tools installed on high performance clients, and high performance hardware on server. The access to SCDT via web browser guarantees the access to a wider range of city users. For example, the SCDTs of Wellington and Shanghai are developed using the Unreal Engine [46]. Even if such solution can provide stunning 3D visualizations and effects, it requires high-end hardware to render the 3D models on a web browser limiting their distribution. Therefore, hereafter was reported a more detailed analysis focusing only on those SCDTs for which a web interface is freely available.

Table 1. Requirements for Smart City Digital Twins

| # | Name | Description | |
|---|---|---|---|
| R1 | Network Modeling and Management | The modeling of data networks including gateways, brokers, devices, services and API, external services as web pages, and protocols. | Field Interoperability |
| R2 | Hierarchical modeling of entities | To model data for terrain, city building and shapes (gardens, roads), services, heatmaps, traffic flows, services, IoT devices, public transportation, etc. To be retrievable with relational, geographical, and temporal queries. | |
| R3 | Logics for data transformation | To transform data collected, for example from IoT sensors, and other sources and transform them into different data models and formats. For example, collecting data from some web service, GIS, FTP and process them for interoperability and for ingestion. | |
| R4 | Smart Data Model compatibility | To guarantee interoperable and replicable Smart Cities, interoperability at level of data formats, FIWARE Smart Data Models, etc. | |
| R5 | Smart City Federation | Model and data federation among platforms at level of protocols and APIs | |
| R6 | Integration with workflow management systems | To enable ticket/event management. For example, when a fault is detected, it is highlighted in the SCDT and linked to a CMMS (Computerized Maintenance Management System). | |
| R7 | Terrain information and elevation | Terrain elevation must be taken into account to properly elevate the city buildings and to model city hills and surrounding mountains | Data and computing for representation |
| R8 | Ground information | Road shapes and names, names of squares and localities, etc., exploiting orthomaps, with eventual real aerial view patterns, and the graph road. | |
| R9 | Heatmaps | To be superimposed (with variable transparency) on the ground level without overlapping the buildings, to represent distribution of temperature, pollutant, noise, humidity, vegetation, etc. | |
| R10 | Paths and areas | To be used to describe perimeters/shapes of gardens, cycling paths, trajectories, borders of gov areas, elements of origin destination matrices, traffic flows, people flows, trajectories, pipes, severs, etc. | |
| R11 | Data analytic | Data analytic processes must be available to let the user develops and/or execute specific data analytics: prediction, traffic flow reconstruction, anomaly detection. | |
| R12 | Single Services | To mark the positions of services, IoT Devices, Point of Interest (POI), Key Performance Indicator (KPI), moving devices as fleets, etc. | |
| R13 | Buildings of the city | Each single building should be represented. Multiple LoD could be included: (i) simple LoD1 structures, or (ii) higher LoD structures represented as 3D meshes, and (iii) BIMs | |
| R14 | Automated 3D building construction | (i) 3D buildings must be created automatically, to be able to scale and replicate the SCDT framework; and (ii) the used software must be released with open or free license. | |
| R15 | Additional 3D entities | To augment the realism of the 3D representation. For example (i) trees, benches, fountains, semaphores, and any other city furniture, and (ii) water bodies to better represent rivers, lakes, etc. | |
| R16 | Dynamic 2D/3D structures | Elements such as PINs, shapes, paths, should be represented in 3D dynamically, changing color and shape according to their kind or some real-time value. | Distribution and interaction |
| R17 | Dynamic data management | To have elements to be automatically reported in the SCDT as soon as they are included in the platform, event driven rendering of data. | |
| R18 | No reloading | Changes in the SCDT must be rendered without the needs of a full reload of the map. | |
| R19 | View Map controls | To change the point of view by zooming, rotating, tilting, and panning the scene. | |
| R20 | Dynamic sky and lighting | To model and show different sky conditions and to change the light source position, simulating different times of day/night. | |
| R21 | Building picking/manipulation | To select single building to: (i) show detailed information, or (ii) move into a BIM view of the building, or (iii) to change the building 3D model. | |
| R22 | Services and element data access | To show data associated with IoT Devices, POI, KPI, shapes, paths, etc., including real time and historical data. | |
| R23 | Independent element management | To hide, show, replace specific elements (e.g., to disable the building view to see only the city PINs, or to load different heatmaps or paths) | |
| R24 | Web player | The SCDT must (i) be accessible thought a web browser without additional plugins, and (ii) the player must be released with open or free license. | |
| R25 | Business logic call-back | To provide the possibility of selecting an element (3D, PIN, ground, heatmap) to provoke a call back into a business logic tool for intelligence activities, analytics, etc. | |
| R26 | Underground and elements inspection | To provide the possibility of selecting and inspecting specific areas and see detailed 3D elements placed underground, such as water pipes, metro lines, etc. | |

Table 2. Comparison of platforms for SCDT. (*) functionality implemented not supported by wide examples.

|  | Helsinki [34] | Rotterdam [36] | Berlin [37] | Stockholm [38] | Zurich [39] | Boston [42] | Snap4City this paper |
|---|---|---|---|---|---|---|---|
| R1 | No | No | No | No | No | No | Yes |
| R2 | Yes (PostGIS) | No | No | No | Yes (Geoportal) | No | Yes (Km4City) |
| R3 | No | No | No | No | No | No | Yes |
| R4 | No | No | No | No | No | No | Yes |
| R5 | No | No | No | No | No | No | Yes |
| R6 | No | No | No | No | No | No | Yes (*) |
| R7 | Yes | No | No | Yes | Yes | Yes | Yes |
| R8 | Yes | Yes | Yes | Yes (fixed) | Yes | Yes (fixed) | Yes |
| R9 | No | No | No | No | No | No | Yes |
| R10 | Yes | Yes | No | Yes | Yes | No | Yes |
| R11 | No | No | No | No | No | No | Yes |
| R12 | Yes | No | No | Yes | No | No | Yes |
| R14.i | Yes | Yes | No | No | Yes | No | Yes |
| R14.ii | Yes (LoD3) | Yes (LoD3) | Yes (LoD2) | Yes (LoD3) | Yes (LoD2) | Yes (LoD2) | Yes (LoD3) |
| R14.iii | No | No | No | No | No | No | Yes |
| R15.i | Yes | Yes | No | Yes | No | No | Yes |
| R15.ii | Non free | Non free | n/a | Non free | Non free | n/a | Yes |
| R16.i | Yes | No | No | Yes | Yes | Yes | Yes |
| R16.ii | No | Yes | No | No | No | Yes | No |
| R17 | Yes | No | No | Yes | No | No | Yes |
| R18 | No | No | No | No | No | No | Yes |
| R19 | Yes | Yes | Yes | Yes | Yes | Yes | Yes |
| R20 | Yes | Yes | Yes | Yes | Yes | Yes | Yes |
| R21 | Yes | Yes | Yes | No | Yes | Yes | Yes |
| R22.i | Yes | Yes | Yes | No | Yes | Yes | Yes |
| R22.ii | No | No | No | No | No | No | Yes |
| R22.iii | No | No | No | No | No | No | Yes |
| R13 | Yes | Yes | No | Yes | No | No | Yes |
| R23 | Yes | Yes | Yes | No | Yes | No | Yes |
| R24.i | Yes | Yes | Yes | Yes | Yes | Yes | Yes |
| R24.ii | Non-free | Free | Free | Non-free | Limited | Non-free | Free |
| R25 | No | No | No | No | No | No | Yes |
| R26 | Yes (*) | Yes (*) | No | No | No | No | No |

In **Table 2**, a comparison among the inspected solutions is provided considering their compliance with the defined requirements.

The SCDT of the city of **Helsinki**, includes a 2D and 3D interface where is it possible to load LoD3 [24] building models together with terrain elevation data (covering R7 and R13.ii). Each building is pickable (R21.i) to access to additional information and paths to delimit specific areas can be visualized (R10). Different orthomaps can be loaded (R8) and PINs are displayed mostly to indicate POIs and services (R12 and R22). However, it seems mostly a static representation since there is not integration with IoT data or other kinds of real-time city related information. We do not find detailed description on the interoperability and even if some simulations are described in [33], they seem not to be integrated in the web interface. The solution proposed by the city of **Rotterdam** exploits multiple LoD building models (R13.i and R13.ii) and additional 3D entities (R15), and do not implement terrain elevation. The ground orthomaps cannot be changed, neither heatmaps nor service's information are available. Similarly, for **Berlin**, the solution includes pickable LoD3 models and the possibility to hide/show single buildings (R13.ii, R21.i). Cast shadows can be visualized and animated according to the time of the day (R20). However, services, paths, heatmaps and real-time data in general are not implemented. The city of **Stockholm** implemented many aspects of the Digital Twin concept, such as services (R12 and R22), LoD1/LoD3 buildings (R13.i and R13.ii), others 3D entities (R15) and dynamic cast shadows (R20). However, the solution lacks in the implementation of heatmaps and real-time sensor readings. The SCDT of **Zurich** implements different versions of LoD2 buildings (R13.ii) that can be selected to show specific information in a popup (R21.i). Additionally, buildings can be shown/hidden on user demand together with building construction plans (highlighted with a different colour). Additional 3D entities (R15) are included -- i.e., trees, with different details that can be selected by the user as well as terrain elevation (R7). Paths are used to delimit areas and to highlight street elements with pickable functionalities (R10). Cast shadows are not implemented and different illumination conditions can be used to show the model at different times of the day (R20). The solution implements also a 3D measuring tool and a pedestrian-view modality. As for the Helsinki case, no details on the interoperability are given, except notes on the usage of Open Government Data. Although in [39] some climate analyses are described, they are not included in the web interface neither as heatmaps nor for simulations. For the city of **Boston**, LoD2 building models (R13.ii) are visualized and classified in three classes: actual building, building under construction and approved projects. Trees are represented (R15) and illumination with cast shadows can be changed by the users (R20). Terrain elevation is considered (R7).

To summarize, the compared solutions providing a web interface are closer to simple 3D city representation models rather than to a complete SCDTs. More specifically, some of them do not address / represent information like heatmaps, services, and paths and none of them seems to be able to aggregates and visualize real-time data, neither to implement analytic processes to perform predictions and simulations. Interoperability requirements seems not to be supported, limiting the possibility to deploy such solutions on different scenarios. To the best of our knowledge, the proposed Snap4City solution is the only freely available SCTD able to model and represent a wide range of data types, show real-time and historic information related to IoT devices and Entities of the web of things, and perform scenario simulations, i.e., What-If analysis, providing to the user a larger set of tools to inspect, assess, change, simulate and study the urban environment. Moreover, by satisfying the interoperability requirements, efficient solutions for data ingestion and representation are available, facilitating the deploy of the SCDT and guaranteeing its continuous update.

*2.2 Article Aims and Structure*

The main contributions of this paper are the following:
- An extended set of requirements, organized in three main groups of needs and their analysis: Field Interoperability, Data and computing for representation, and Distribution and interaction. The requirements resulted to be much wider and precise with respect to those of [24] and [27].
- A wide State of the Art analysis and comparison that takes into consideration the most relevant SCDT solutions with respect to the identified requirements.
- The novel high performance data distribution and rendering engine of the SCDT which permits the access to complex 3D SCDT representations from regular browsers. The new modelling and distribution engine is based on the concept of *FusionLayer* that offers better performance reducing server requests and lowering client resource usage. It substituted the former data layered structure of the interactive web interface [27].
- The introduction of a wider set of new functionalities never presented before for Snap4City: (i) individual 3D building model selection and substitution; (ii) BIM (Building Information Modelling) management; (iii) road graph as interactive elements; (iv) traffic flow density in real time as animated arrows.
- A more in deep description of the SCDT engine and functionalities offered by Snap4City.

In Section 3 the SCDT general architecture is reported together with descriptions on data modelling, focusing on semantic modelling and Snap4City/Km4City ontology. Section 4 presents data and processes (used to obtain reconstructions, predictions, and simulations) for the corresponding representations and usage in the SCDT. Sections 5 is devoted to illustrating SCDT distribution through an open interactive the web application implementing the rendering engine. In Section 6, the case of study of the city of Florence in Italy is presented. Finally, conclusions are drawn in Section 7.

## 3. General Architecture and Data Modelling

The presented SCDT is built by exploiting the Snap4City platform [47, 48] (www.snap4city.org). Snap4City is an open-source IoT platform coordinated by the DISIT Lab of the University of Florence. The architecture of Snap4City is reported in **Figure 1**, in which the areas mainly addressing the above-mentioned groups of requirements are highlighted. The **Field Interoperability** and data processing area provides the capabilities of collecting data from any sources and exchanging data in push toward any brokers, gateway, and services.

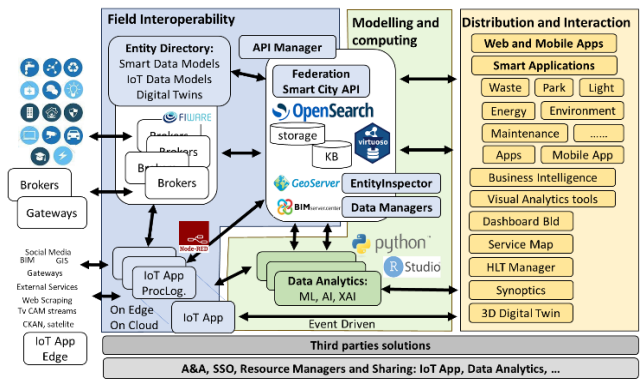

Figure 1. Snap4City Architecture overview.

This area is interoperable with a large number of protocols and formats (https://www.snap4city.org/65), compiling with R1, and enabling federation of Smart Cities (R5) [49]. Real-time data, as well as event driven streams, are ingested using IoT Brokers and IoT App to be indexed and shadow stored into an OpenSearch cluster. Thus, they are accessible for other consumer processes and brokers. The internal brokers are based on Orion Broker NGSI (also compatible with smart data models, R4) for the data retrieved in push, or event driven. IoT App/Proc.Logic are processes in Node.js, and they are used for the data interoperability with third party services such as: GIS (Geographic Information Systems), ITS (Intelligence Transport Systems), TV cam services, CKAN open data networks, BIM servers, social media, data gateways, etc. The Node.js processes are created via Node-RED with Snap4City set of libraries and microservices and are used for data transformation (R3) [30]. The collected data are stored into a set of storages (R2): RDF Virtuoso for the graph semantic relationships (spatial, temporal, and relational) among the digital twin entities according to Km4City ontology [50], Open Search for time series data, specific Geoservers and databases for Heatmaps, traffic flow, origin destination matrices, orthomaps, etc.

The road graph is represented in the Km4City ontology with entities to describe *Roads*, *RoadElements*, *Lanes*, together with possible *Restrictions* (see **Figure 2** for a simplified representation) retrieved from OpenStreetMap. Sensors are mapped into the Km4City knowledge base using Smart Data Models of FIWARE [51] or other standards, as well as from IoT Device Models of Snap4City. These include traffic, weather, and air quality sensors as well as other kinds of sensors used to model sharing vehicles, bike racks and parking occupancies, people counting, etc. POIs (Point of Interests) are modelled as *Service* entities specialized in more than 20 categories (e.g., Accommodation, CulturalActivity, Emergency, Financial, Commercial, Cults), and 500 subcategories. Heatmaps, that can describe dispersion of pollutant, temperature, energy, humidity, traffic/people density, etc., are managed by the Snap4City GeoServer, a GIS Server able to serve georeferenced data in tile with user-defined dimensions using the WMS/WFS protocols over HTTPs.

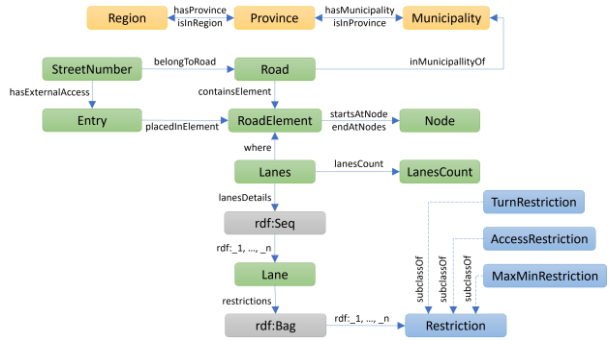

Figure 2: Road Graph Data Model (partial) as represented in the KM4City knowledge base.

Building 3D models (see Section 4.1 for details on 3D model construction) are stored exploiting a tiled approach following a hierarchical Z/X/Y folder organization used by most of GIS applications like OSM (Open Street Map), etc., where Z is the zoom factor (fixed at 18 for our tiles) that describe the tile dimension, while X and Y are the tile coordinates. This solution was adopted to guarantee a fast model loading through the browser interactive web interface (see Section 5). The complete city map was divided in non-overlapping tiles and each building was uniquely associated to a single tile considering the centroid of the building shape at ground. Note that, even if the models are grouped in tiles, each single building is represented as a separate entity in order to enable the picking functionality.

Finally, the openMAINT Computerized Maintenance Management System, is integrated into Snap4City to handle ticket management for maintenance and operation (R6).

Indeed, the exploitation of the Snap4City IoT platform has been a key factor to enable a fast and reliable data acquisition and management, and, using dedicated queries and API, to retrieve any kind of data and display them on the digital twin interactive web interface, to respond to a set of requirements on interoperability (R1 – R6).

Data & Computing for Representation and Distribution & Interaction layers are the most challenging aspects in the development of a SCDT and for this reason they are described in the following sections.

## 4. Data and Computing for Representation

The **Data and Computing** group of requirements describes which data need to be addressed in the SCDT and the related computing capabilities to produce them. For example, to obtain traffic flow reconstruction, 3D reconstruction, data time series predictions, and simulations. In this paper, the focus is on the computational needed for the production of the 3D models.

At first, a three-dimensional representation of the city is included in order to build an accurate replica of the urban structures. This encompasses the terrain mesh, the building models, and additional 3D entities to augment the realism of the representation. **Terrain elevation** is represented by exploiting the RGB (Red Green Blue) encoded DTM (Digital Terrain Model) retrieved from the GeoServer (see Section 4.1) that is used to generate the mesh of the terrain using the Martini tessellation algorithm (R7). A tiled organization must be exploited to load only the terrain areas on the basis of the user's view in the browser. Additionally, multiple resolutions are used to model with high precision the urban area and reduce the burden to represent city surroundings using lower resolution representations. Note that, all the following entities that are included in the proposed SCDT must be elevated accordingly to the terrain / TDM to avoid sunken or floating elements. Then, **3D building models** with different LoD are provided. Simple LoD1 representations (R13.i) and more accurate LoD3 models (R13.ii) are included and can be selected at run-time by the user. BIMs (R13.iii) are accessible from the 3D map using a dedicated interface to perform more in deep inspections in all the construction details of the building. Dealing with a huge number of buildings, as it is required to represent major cities, is undoubtfully a challenge, in particular when the system must be accessible by any device, without particular hardware requirements, especially for client side. For this reason, specific solutions must be devised. In our case, buildings are organized in tiles at a fixed zoom level and a set of new functions/algorithms have been developed are used to exploit a tiled representation at multiple zoom levels (see **Section 5.3**). Alternatively, multi-zoom grouping could have been exploited; however, such a solution would have required to replicate the models in the digital twin storage (to precompute aggregations at different zoom levels) increasing the memory occupancy and reducing the maintainability since changes should have been replicated in all the different groups of tiles. **Additional 3D entities** (R15.i), such as the **trees** (or other city furniture) may be included in the map using 3D models positioned according to the information provided by the municipality. Other entities can be added as well, for example, 3D models of **airplanes** have been included in the airport area. Appropriate model loading procedure must be used to represent additional entities. Differently from the buildings, a limited number of models are required to represent certain kind of entities which are replicated in a large number of positions (e.g., trees belong to 3-4 categories which are replicated hundreds of times with different scale/sizes). In any cases, due to the high number of different models, an efficient solution must be used to retrieve 3D models from the server for each specific model, and in some cases, there are models that need to be replicated in multiple locations.

The terrain texture is created by merging multiple images: the orthomap of the terrain, over which different heatmaps can be shown on user demand. Different **orthomaps** (R8) can be required to be shown by the users and need to be loaded at run-time on user demand to display specific ground level information (e.g., road map or satellite images). They also have to follow the TDM. **Heatmaps** (R9) are usually semi-transparent superimposed on the orthomap, and can be activated by the user to visualize distributions of temperature, pollutant, noise, humidity, shops, vegetation, etc. Each heatmap is loaded with its corresponding colour-map. Heatmaps can be static or animated: static heatmaps are provided as single PNG images, while animated ones are provided in GIF format with multiple images rendered sequentially with a custom delay. Since heatmaps must be superimposed on the ground orthomap and TDM, the user must have the possibility to set the heatmap opacity. This functionality can be achieved using a multiple image merging process to be able to obtain a single terrain texture that mix orthomaps and heatmaps with different level of opacity (see Section 5.4). **Services** are loaded from the Km4City knowledge base and represented as PINs to show positions and data of IoT devices, POIs, as for example bus stops (R12), parking, banks, accessible floating bikes, etc. Considering the huge number of services that can be scattered over the city area, efficient functions are required to retrieve and show the associated PINs. **Cycling paths and roads,** stored in the KM4City knowledge base, are represented as line segments (R10), and are selectable to get more information such as: velocity, category, vehicles kind, status, size, etc.

The **Computing** processes provide capabilities of developing and put in execution a large range of Data Analytics by using ML (machine learning), AI (artificial intelligence), simulations, optimisations, etc. (R11), e.g., [52, 53]. The data analytic processes are used for: (i) **operative computing** to obtain predictions, key performance indicators, making strategies, and for calculating heatmaps, origin destination matrices, traffic flow distributions, that are information rarely available from external sources; (ii) **generative computing** for the construction of the digital twin representation in 3D. In **Figure 3,** a schematic representation of data and computing processes is reported.

Given the complexity of the creation of the 3D models and their relevance in the context of SCDTs, hereafter additional details are reported.

### 4.1 Constructing 3D Models

As can be seen from the block-diagram reported in **Figure 3**, the 3D Map construction process starts by gathering different kinds of input data that are elaborated by several sub-processing blocks and finally indexed/archived in specific storages on the Snap4City platform. Inputs include: (i) street-level and aerial (i.e., orthomaps) RGB photos to obtain roof and façade textures and to model possible High Value Buildings (HVB); (ii) building plant shapes from OSM or municipality cadastre (used to geographically localize the buildings); (iii) building height information (in the format of GEOJson files) and/or Digital Surface Model (DSM) data to properly model the 3D structures; (iv) DTM data to compute a terrain level and to put the buildings and other entities at the right elevation and perspective. Moreover, BIM and additional 3D entities are considered, which may require to be obtained via some format conversion. Hereafter, more details on the sub-processing blocks are reported.

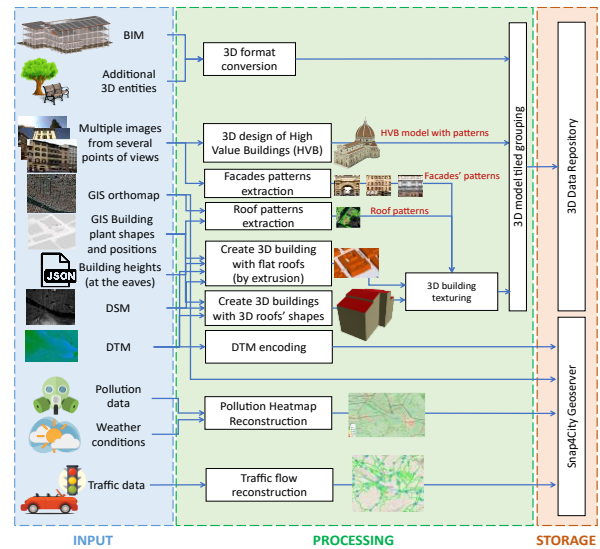

Figure 3. Generative and operative computing process to create the 3D structures and heatmaps.

**3D Design of High Value Buildings.** To obtain high quality models for the HVBs manual 3D design or automatic computer vision techniques, such as Structure from Motion, can be employed. The obtained models are then put in the right scale, position, and elevation (w.r.t. the DTM).

**Façade and roof pattern extraction.** Regarding the roof and façade patterns, they are respectively extracted from orthomaps and street level images. In order to obtain an accurate orthomap segmentation to extract the roof texture, a deep net was used [54] to find the similarity transformation required to locally warp the orthomaps and make them accurately fit the building plant shapes [55, 56]. Diversely, façade's patterns are extracted by segmenting the building façade and then rectifying them using planar homographies. Both roof and façade texture are then applied on the 3D building models using Blender plugins developed for these specific tasks.

**3D Building construction.** 3D structures of ordinary buildings can be obtained with two different approaches. Flat-roof buildings are obtained by extrusion from the building shapes at the ground level taking into account the building height. Such height attribute can be obtained from manual measurements of the eave heights, or by evaluating the average height of the DSM samples included into the building plant shape. Differently, 3D-roof buildings are obtained by analysing the DSM and fitting on its samples planar primitives to describe the different roof slopes. Such a process includes spatial clustering [57, 58], multiple linear and planar robust model regression [59]. Both flat-roof and 3D-roof building models can be put at the right terrain elevation exploiting the information encoded in the DTM. The building construction process is carried out automatically (R14.i) and code is freely available on DISIT GitHub[*] repository (R14.ii) for Snap4City.

Additional 3D entities (R15.i) such as trees, streetlamps or other minor urban structures/furniture can be obtained from accessible 3D repositories and placed into the map exploiting positioning information and size, which can be obtained from Open Data as well as from the municipality.

**High resolution DTM encoding.** To exploit the DTM as a terrain level in Snap4City interactive user interface, the DTM, expressed in float values, has been converted to the RGB format and deployed in the Snap4City Geoserver. In order to accomplish the DTM conversion, we use the following mapping function:

$$\begin{cases} R = \left\lfloor \dfrac{100000 + 10v}{256^2} \right\rfloor \\ G = \left\lfloor \dfrac{100000 + 10v}{256} \right\rfloor - 256R \\ B = \lfloor 100000 + 10v \rfloor - 256^2 R - 256 G \end{cases}$$

where $v \in \mathbb{R}$ is the DTM raw value. In this way we can obtain an RGB image able to encode elevation differences up to 0.1m. Then, the encoded DTM is loaded into the Snap4City GeoServer to be retrieved in real-time by a scalable and tiled approach.

## 5. Digital Twin Distribution and Interaction

In this section, the complexity of addressing the representation of the whole SCDT (3D models, data, simulations, etc.) is discussed. The first subsections are devoted to the general view, and to describe the Dynamic 3D representations and interactions, and the Hierarchical layered structure. Finally, the new **Fusion Layer** approach and algorithm is described in which all the entities are efficiently managed for distribution and rendering with interactivity. The new approach also constrained to integrate different strategies for: Terrain, orthomaps, and heatmaps layers (see Section 5.4); and for supporting interactivity in What-If analysis (see section 5.5).

The framework developed for distributing the SCDT as a 3D multi-data map dashboard in the Snap4City platform is able to retrieve and reassemble all the data specified by the requirements from R7 to R15, including: different version of 3D models of buildings (LoD1, LoD3, BIM) and additional entities, heatmaps (for traffic flow, pollutant dispersion, etc.), services as PINs (IoT, POI, etc.), 3D terrain from DTM, sky pattern, ground information, paths and areas. The framework implements a client-side business logic that exploits a series of REST API (smart city API) calls to load the data independently on user demand (as requested by R23). For example, the 3D building tiled representations are retrieved via HTTPS protocol, as well as data for POI, IoT devices / time series, paths, etc. obtained with specific geographic queries on the SuperServiceMap of Snap4City, a GIS interface of the knowledge base. Differently, heatmaps (static or animated), orthomaps, and the encoded DTM are retrieved via WMS protocol over HTTPS by querying the GeoServer, limited to the portion of the map visualized by the user. Moreover, collected static and real-time data, semantically indexed in the Km4City graph-based RDF Knowledge Base ontology, can be retrieved using microservices and/or Smart City APIs [50] based on secure HTTP REST calls to execute spatial, temporal, and relational queries toward the SuperServiceMap. In this way, every time a new element is indexed into the knowledge base it can be automatically displayed on the SCDT interface (R17).

The rendered solution has been implemented via layered WebGL and APIs, in order to access to the GPU thanks to the passthrough available in web browsers without the needs of plugin[†], therefore satisfying R24. Exploiting the open-source Deck.gl library, the web application has been realized by using a custom implementation and management of the *ViewState* object, in which the geographical information for the map (such as latitude, longitude, zoom, etc.), are defined. Taking into account the viewing position and angle, elements are loaded from the nearest to the farthest with respect to the point of view. This dynamic loading reduces the amount of resources needed to display the current scene, since elements outside the ViewState are not processed. This is particularly useful when dealing with Smart City covering wide geographical areas.

Finally, it is worth noticing that all the displayed elements are handled independently and can be shown or hidden at run-time on users' demand (R23) without requiring a full reload of the map (R18). To achieve this, a visualization solution has been designed to be capable to work in a layered way to have different elements represented in specific layers. Finally, resulting 3D representation can be panned, rotated, and zoomed by the user using the mouse or the in-map buttons (R19).

### 5.1 Dynamic 3D representations and interactions

To better show particular information on a 3D map, new kinds of representations have been designed and developed. For instance, traffic flows, typically represented by coloured lines over the ground, can be better represented using static or dynamic 3D elements (R16). In the SCDT, traffic flow densities are shown as raised crests (with amplitude proportional to the computed traffic density values) and coloured using a colour map that can be defined by the user.

---

[*] https://github.com/disit/3d-building-modelling

[†] https://github.com/disit/dashboard-builder

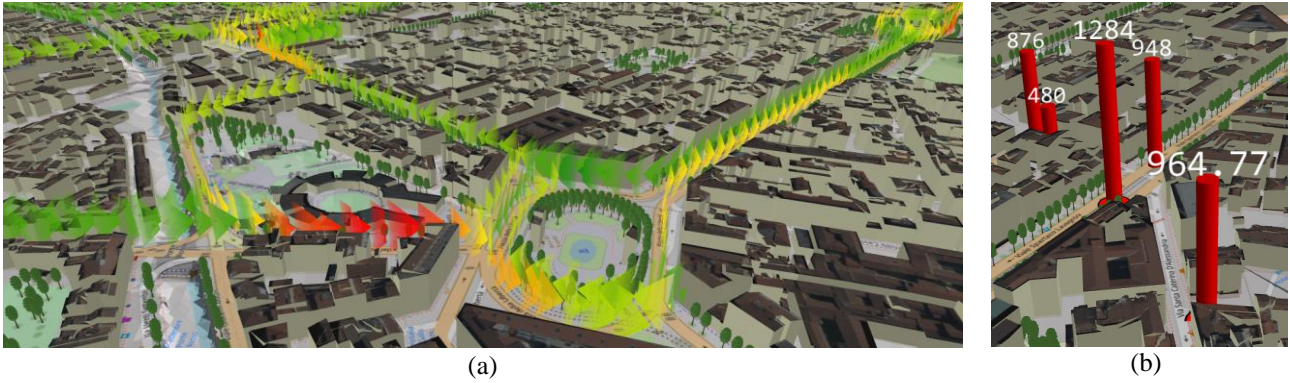

(a)                                         (b)

Figure 4. Example of dynamic 3D elements: (a) Traffic flow density represented as animated 3D arrows. (b) 3D column representing traffic sensors.

Alternatively, traffic can be displayed as animated arrows (see **Figure 4a**) moving with a speed related to the real-time traffic density (slower for congested roads, faster for free roads). Both the crests and the arrows must follow the terrain elevation, and by hovering the mouse over the elements pop-ups are displayed reporting traffic information. Due to the nature of the problem, traffic flow polylines can be fragmented. Then, for each segment the points are passed to the GPU and interpolated to render the crest profile exploiting the fragment shader with a considerable speed-up. Additionally, to animate the arrows different animation times must be used to represent different traffic conditions: faster for free roads, slower for congested segments. Similarly, 3D columns (see **Figure 4b**) are used to represent real-time data values registered from IoT sensors: the value of any sensors can be shown with a 3D cylinder (or other solid) shape, positioned where the sensor is, and whose height is proportional to the considered metric, while the actual value is reported in textual form on top of the cylinder. The colour of the column represents the category of the sensors, and all these elements can be customized by the user.

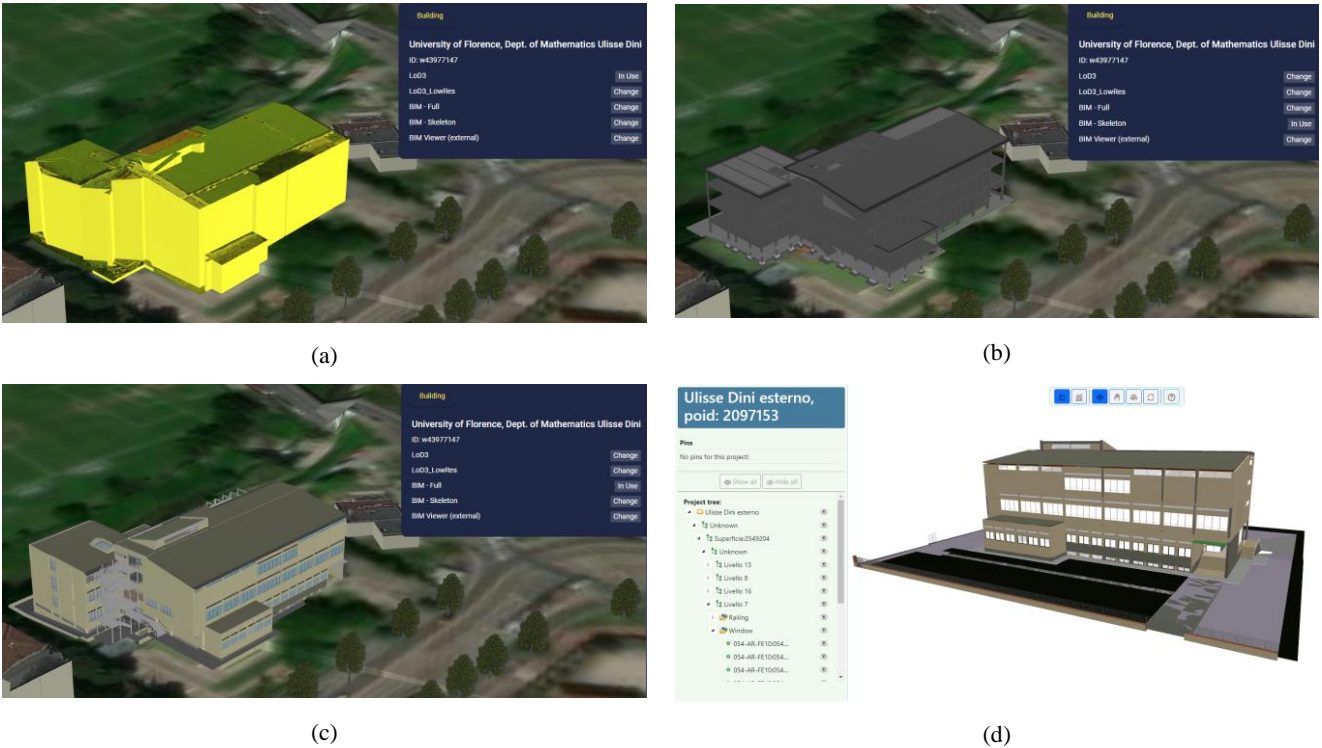

Figure 5: Example of 3D building substitution and BIM integration. In (a) the default LoD3 building model is selected (highlighted in yellow) and a panel in the top left corner of the interface show building related information and other available 3D models. By clicking on the buttons, different models are loaded in the SCDT web interface without requiring any refresh. For instance, in (b) a skeleton structure of the building is shown, while in (c) a GLB model obtained by converting the IFC BIM representation is displayed. Finally, in (d) the full BIM of the building is presented: such representation is accessible from one of the panel buttons and opened in a dedicated dashboard with additional functionalities for BIM inspection.

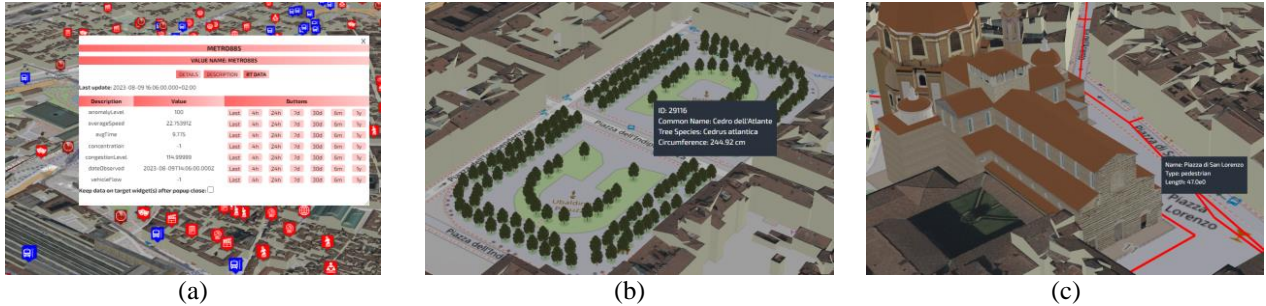

(a)                  (b)                  (c)

Figure 6. Example of interactive pop-up with specific element information for services as PINs (a), additional 3D elements, as trees, (b), and road elements (c).

Additional interactivity is provided for buildings, additional 3D elements, PINs, and paths. Each building model can be selected (R21.i) to access to additional information that are reported on a panel in the map. To achieve such functionality, an appropriate in-tile indexing is provided to single out a 3D building from any tile. From the building panel, the user can select different 3D representations of the same structure (R21.iii) to change at runtime the model used in the map, and access to a BIM representation (R21.ii) that is visualized and that can be inspected in a dedicated interface. In **Figure 5** an example of building substitution is reported. Such a functionality is useful to modify the buildings included into the map, for example to appreciate different possible city construction plans, to be showed to wide public, improving the citizen engagement toward the urban landscape evolution.

Information regarding additional 3D entities (e.g., trees), road elements, and cycling paths can be accessed by the user by hovering the mouse on the specific element so to bring up pop-ups with additional details (like the street name, the tree species, etc.). Services are represented as 3D PINs with a different icon and colour according to their semantic category. By selecting them a pop-up window is shown reporting real-time or historic data (R22) – see **Figure 6**. Moreover, by clicking on the user interface events can be raised to provoke a call-back to business logic (R25) and to show additional data in separated widgets included in the same dashboard (e.g., time trends) requiring client-side business-logic functionalities [60] to let different widgets exchanging information.

Scene illumination (R20) has been modelled with two types of lights: an ambient light to affect all the scene, and a directional light to model the position of the sun (calculated according to [61]). This process creates the lights and shadows of the scene, and it is useful to simulate when a particular area is well illuminated or not.

### 5.2 Hierarchical layered structure

In the previous version of Snap4City SCDT [27] the tile loading was based on a hierarchical layered structure mainly exploiting the TileLayer of Deck.gl with some modifications to handle specific data needs. For example, to display the buildings at different zoom levels, we modified the SceneGraphLayer, nested in the TileLayer, to handle multiple different 3D models and reduce web traffic and server calls, changing the default Deck.gl behaviour. However, such a solution gives rise to problems when dealing with a huge number of tiles. Since some of the elements to be displayed are organized at a fixed zoom level, when showing wide areas, a large number of tiles must be loaded in the client GPU and CPU, augmenting the use of resources and worsen user experience due to lags and general slowness. Differently, in the previous version, 3D crests and columns, PINs, and paths, were loaded without using a tiled approach to obtain all the entities in a single request regardless of the zoom factor. Indeed, such elements are always the same at any zoom level (differently from heatmaps or orthomaps that could be provided with finer or rougher details according to the zoom). This was performed to avoid additional server calls. However, such a solution produced a significant slowdown in rendering the data since only after completing the download globally the entities were drawn. Note also that, in the former version, only the elements in the viewport were loaded from the server. Moving to a different area would have required to make new server requests and render the elements in the map, retrieving anew some entities previously downloaded. In order to solve these issues, the novel *FusionLayer* was developed and hereafter described. A graphical representation of the novel layer structure is shown in **Figure 7**.

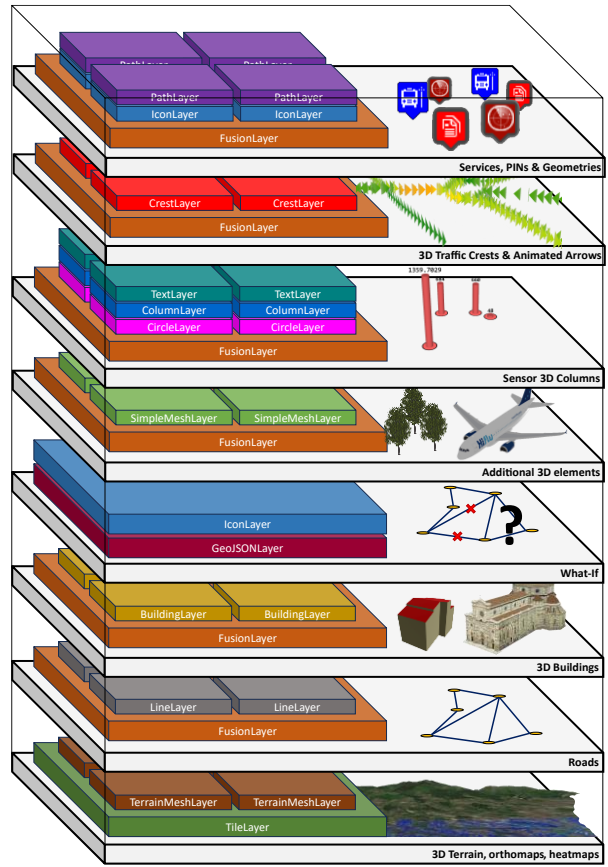

Figure 7: layered architecture used to develop the interactive web-based interface to represent all the data required in the SCDT exploiting the novel *FusionLayer*.

*5.3 FusionLayer*

To keep retro-compatibility and exploits some native Deck.gl functions, the *FusionLayer* was derived from the *TileLayer* introducing additional functionalities:

- Reduce server calls by introducing an element cache management on the client exploited by two main functions *bottomUpFusion* and *topDownFusion*.
- Avoid excessive use of client resources by implementing the *deepLoadFunction* to retrieve elements in tiles at a given zoom and then group the results at run-time in bigger tiles at arbitrary zoom levels.

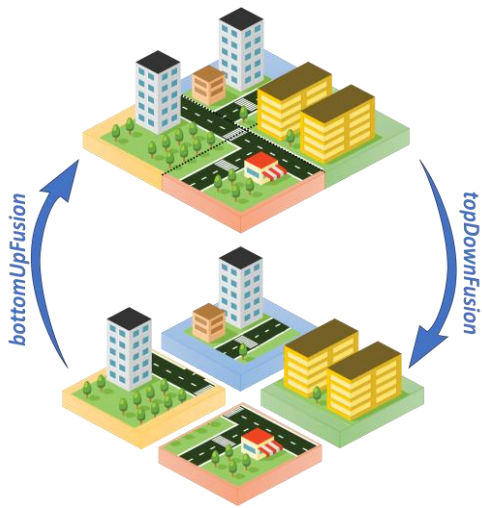

Figure 8: *bottomUpFusion* and *topDownFusion*.

The *bottomUpFusion* and *topDownFusion* are two specular functions used to reuse already retrieved data and avoid new server calls, see **Figure 8**. More in details, suppose to start with a tiled visualization at zoom level $z_i$ and then move to a zoom level $z_j$, with $j > i$, i.e., the user is zooming into the map. The *topDownFusion* is used. Exploiting the hierarchical tile organization, the system is able to know which tiles at zoom $z_j$ corresponds to a tile at zoom $z_i$. For each child tile $t_c$ at $z_j$, the system search for a parent tile $t_p$ at $z_i$ and retrieve the cached elements. Then each element position is checked and if is assessed that it belongs to the tile $t_c$, the element is added to the feature array of tile $t_c$, avoiding retrieving again the elements from the server. Similarly, when the zoom level changes from $z_j$ to $z_i$, i.e., the user is zooming out, the *bottomUpFusion* is invoked. In this case, all the elements in the chid tiles $t_c$ at zoom $z_j$ are added to the feature array of the parent tile $t_p$ at $z_i$. For both fusions, if for a tile no data are found in the cache, i.e., the child/parent tile was not previously loaded, then new server calls are executed to retrieve the data, for instance, when the user move to different map areas originally not in view. It is worth noting that, both the *topDownFusion* and the *bottomUpFusion* can work even if the considered zoom levels are not contiguous (i.e., $j - i > 1$). In such a case, the *jumpZoom* functionality is exploited to put in relation child and parent tiles of not contiguous zoom levels.

Some elements to be displayed in the web interface are available at a specific zoom level. For example, building models are grouped in tiles at fixed zoom level 18. Using a naïve solution, if the scene in view covers a wide area, a large number of tiles would been loaded in the GPU slowing down the system performance. To solve this problem, the *deepLoadFunction* was developed. Let us now suppose to have elements available only at zoom level $z_j$ and being in a *ViewState* condition with zoom level $z_i$, with $i < j$. The *deepLoadFunction* initially splits the tiles at zoom $z_i$ in $n$ virtual sub-tile at zoom $z_j$. For each of these sub-tiles the required elements are retrieved with $n$ independent asynchronous requests. Then, when all the responses are collected, the system aggregates all the data and displays them in a unique layer at zoom $z_i$. Note that, data retrieval can exploit also the cached element using the *bottomUpFusion* or *topDownFusion*.

The *FusionLayer* can be specialized using some parameters to work with different kinds of geo-localized data. For instance, at a given moment, it is possible to define a *JsonFusionLayer* to work with plain JSON file, and a *GeoJsonFusionLayer* to deal with GeoJSON data. Nevertheless, the *FusionLayer* can be easily extended to work with additional data formats, such as SVG, CSV etc.

In the actual implementation, the *FusionLayer* is employed to load buildings, additional 3D elements (e.g., trees), paths and shapes, PINs for IoT Sensors, POIs, services, etc., dynamic 3D elements like traffic crests or animated arrows and 3D columns (see Figure 8). For each data kind, a new instance of the *FusionLayer* is instantiated to load elements independently and with different safe context. In particular, using the *FusionLayer*, buildings are always retrieved at tiles with max zoom 18 and displayed at different zoom levels without excessive GPU loads exploiting the *deepLoadFunction*. PINs, paths, 3D crest, arrows and columns, are now loaded in a tiled and dynamic modality, enhancing the interface responsiveness while still limiting the server requests, since already downloaded data could be reused thanks to the fusion mechanism. On the other hand, using this solution, elements that fall out of the viewport are deleted from the client memory after a given delay reducing the computational burden. Moreover, using this tiled approach allows the system to perform a more accurate data retrieval. Indeed, most GIS servers can only provide data included in a squared bounding box, that is adequate when working on 2D maps. Conversely, a perspective view of a 3D map typically draws a trapezoid on the main plane. Therefore, to query all the elements in the viewport a bounding box wider of the viewed area is required, especially when the perspective includes to the horizon. Using our new tiled approach, the system can query the server using finer bounding boxes, so to avoid retrieving elements outside of the viewport, but also being able to query bounding boxes with different size (i.e., different zoom level), smaller for tiles closer to the viewpoint, wider for those that are farther away.

### 5.4 Terrain, orthomaps, and heatmaps layers

Not all the data to be visualized in the interactive web interface of the SCDT need to exploit the novel *FusionLayer*. Terrain data, orthomaps, and heatmaps are provided by the GeoServer in tile of user-specified zoom level and at different resolutions (higher resolutions for higher zooms covering smaller areas and vice versa). In this case the *topDownFusion* and the *bottomUpFusion* cannot be used since for different levels of zoom new data must be retrieved to obtain the correct resolution. Similarly, there is no need to use the *deepLoadFunction* since data can be requested at different zoom levels.

The implemented solution is able to handle a three-dimensional terrain with accurate elevations using the default *TileLayer* to display in tiles the RGB encoded DTM retrieved from the GeoServer. The obtained DTM is used to generate the mesh of the terrain using the Martini tessellation algorithm. Note that, it is possible to mix multiple DTM files, for example with different resolutions: in such a case one of the DTMs has higher priority over the others.

The terrain texture is instead created by merging multiple images: the base image is the orthomap of the terrain (that can be chosen by the user), over which different heatmaps can be shown on user demand. This data integration forms a layer called *TerrainMeshLayer*. The texture merging process is carried out directly inside the GPU to have maximum performance. To handle the selected opacity level, the following equations were used to merge different textures inside the fragment shader:

$$\begin{cases} mix_\alpha = 1 - (1 - \alpha_2) * (1 - \alpha_1) \\ mix_R = \left(\frac{R_2 * \alpha_2}{mix_\alpha}\right) + \left(\frac{R_1 * \alpha_1 * (1 - \alpha_2)}{mix_\alpha}\right) \\ mix_G = \left(\frac{G_2 * \alpha_2}{mix_\alpha}\right) + \left(\frac{G_1 * \alpha_1 * (1 - \alpha_2)}{mix_\alpha}\right) \\ mix_B = \left(\frac{B_2 * \alpha_2}{mix_\alpha}\right) + \left(\frac{B_1 * \alpha_1 * (1 - \alpha_2)}{mix_\alpha}\right) \end{cases}$$

where $\alpha_i$ is the alpha channel of the background image ($i = 1$), and the additive image ($i = 2$) to be merged, while $R_i$, $C_i$, and $B_i$ are the RGB channels. When merging three or more images, the process is performed progressively for each pair, cumulating the next on the first pair merged.

A custom rendering system was implemented in order to add the *SkyBox* feature based on a direct access to the WebGL context, required to include a sky representation into the 3D map.

### 5.5 Managing What-If layer

A key functionality for urban planning and management that can be offered by a SCDT is the possibility to perform simulation and observe the effects produced by a change in the contextual environment modelled in terms of semantic or GIS modelling/representation of the road graph. For example, to observe how vehicle routing can change due to a scenario in which an area is blocked to traffic. Such a functionality is called What-If analysis [62]. This can be applied to different kinds of simulations to understand the impact of scenarios on traffic flow, possible routing approaches, pollutant diffusions, people flow, etc. To achieve this, analytic processes (R11) – the same or similar processes used for the operative computing (see Section 4) – are exploited through business-logic call-backs (R25) to send user input and obtain the results to be visualized at run-time without requiring neither explicit change on the road graph nor map reloading. What-If analysis results and the selected scenario, both represented as elevated lines to be better visualized in a 3D map, are shown to the user using the *WhatIfLayer*, developed from scratch. The *WhatIfLayer* is a composite layer that is formed by GeoJSONLayers and IconLayers that are rendered dynamically as the user interact with the map.

## 6. Case of Study: the SCDT of Florence

In order to validate the proposed Snap4City SCDT framework, we selected as case of study the city of Florence in Italy. Our SCDT encompass the full Florence municipality (as shown in **Figure 9**) covering an approximate extension of 151 km$^2$ (wider than the area of the Florence municipality of 102 km$^2$). Snap4City SCDT of Florence is freely accessible through a web interface[*].

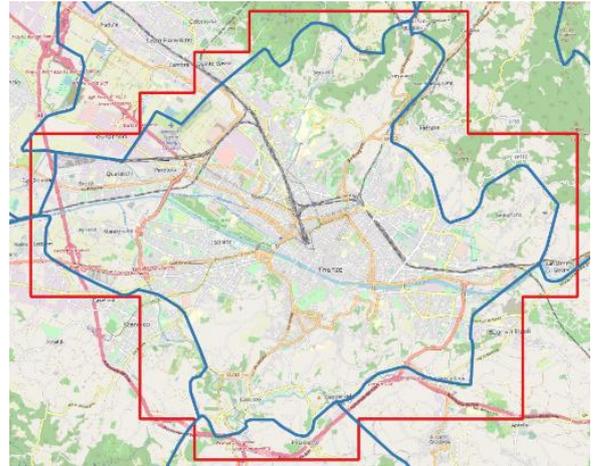

Figure 9. Extension of the modelled Florence area. In blue the administrative border of Florence municipality; in red the area covered by our Digital Twin. This map is shown in the EPGS:3003 coordinate system.

---

[*] See https://digitaltwin.snap4city.org

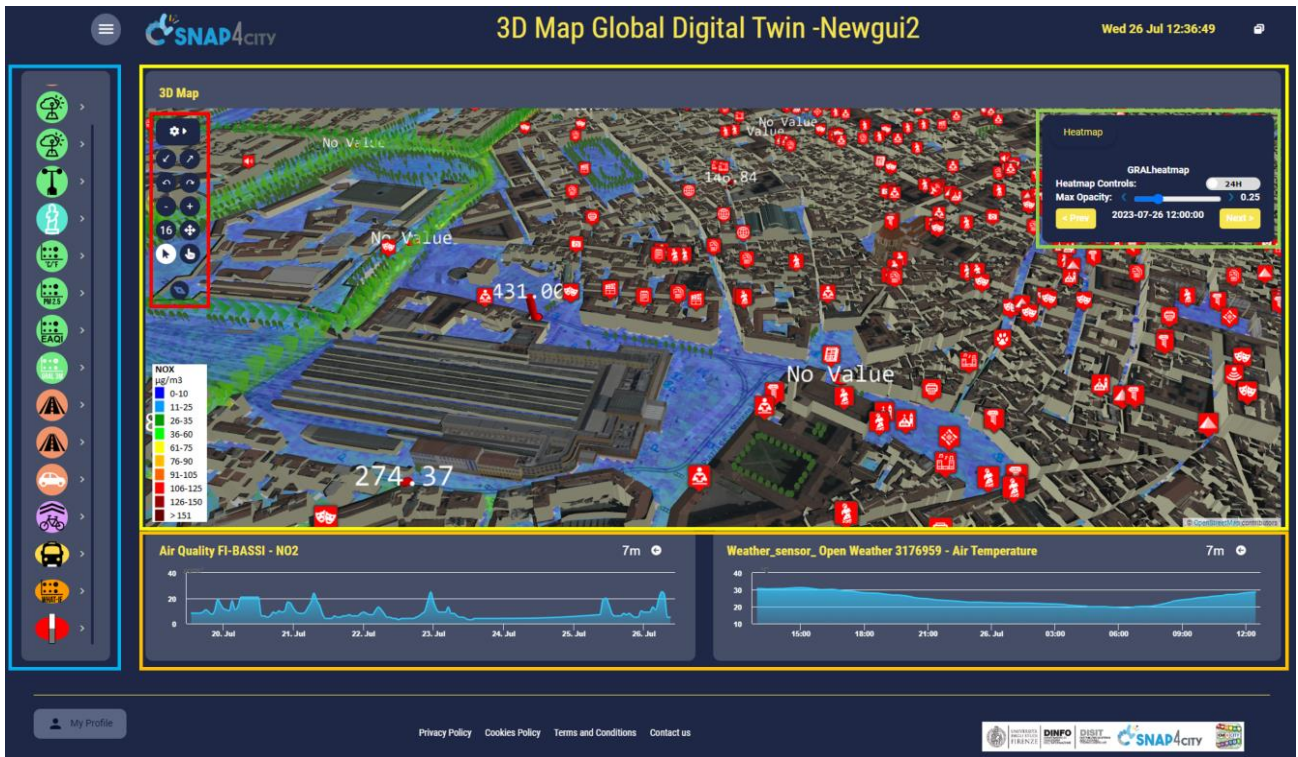

(a)

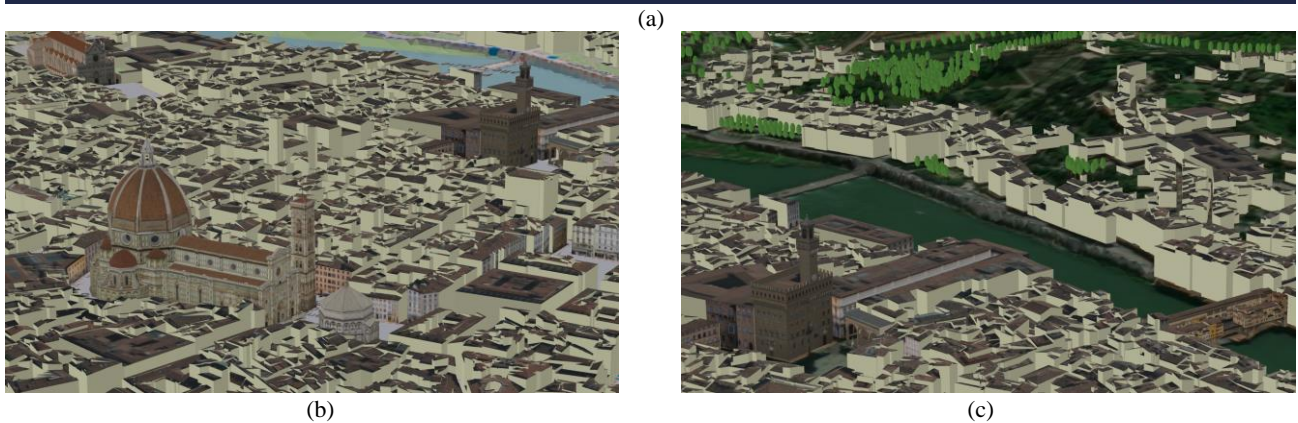

(b)                                                                                      (c)

Figure. 10. Snap4City dashboard showing the Smart City Digital Twin of Florence. In (a) the dashboard is presented with LoD3 and HVB models shown together with PINs, 3D animated arrows for traffic, heatmaps, 3D Cylinders, trees. The blue rectangle highlitgh the selecto menu. In yellow the main 3D multi-data map, with the in-map menu (red rectablge) and the interactive pannel (green rectangle). Under the map, in the orange rectangle, addional widget can be visualized. In (b) a close-up view of the Florence city center. In (c) another close-up showing entities correclty elevated according to the 3D terrain (textured with satellite orthomaps). To try our SCDT of Florence the reader is invited to visit the following link https://digitaltwin.snap4city.org.

In **Figure 10** a screenshot of the interactive web interface of the SCDT, implemented as a 3D multi-data map dashboard in the Snap4City platform, is reported. On the left side, a selector menu is displayed to show/hide information coming from analytic service or real-time data. The main portion of the web interface is devoted to represent the 3D view. In the top-left corner in-map controls are included; in particular the pointing-hand icons can be clicked to activate the picking functionality on buildings shown in the map to access to detailed information and controls to dynamically change the displayed 3D model or access to BIM representation when available. Moreover, the gear icon can be used to access to an in-map menu where the user has the possibility to select different orthomaps, buildings, and activate animations, decorations, and road information from a dedicated setting panel. In the top-right corner a panel can be shown to provide the user additional information and controls. Additional widgets to show time series, histograms, etc., can be included in the same dashboard and updated in real-time using event-driven callbacks to a client-side business logic [63].

Snap4City SCDT includes LoD1 and LoD3 3D building models, and terrain elevation. The DSM and DTM data, modelling respectively the buildings and the terrain, were kindly provided by the "Sistema Informativo Territoriale ed Ambientale" of Tuscany Region. They were obtained from a LiDAR survey and are composed by several tiles covering the city of Florence, with a resolution of 1 square me-ter. 3D building models were enhanced with roof textures obtained from orthomaps of the city of Florence. The RGB photos are tiles with a resolution of 8200x6200 pixels, with partial overlap and rough geo-localization in the EPSG 3003 (Monte Mario / Italy zone 1) coordinate system. The SCDT includes services represented as PINs indicating positions of POIs, IoT sensors, bus stops, parking, pollutant, lights, etc. Thanks to the semantic indexing of data offered by the Snap4City Knowledge Base, different PINs can be represented with specific icons according to their semantic category [64, 65]. Information associated with a specific service can be accessed by simply clicking on the respective PIN: a popup is shown to the user presenting static attributes and, when available, real-time and historical data. 3D column representations are also available to display city traffic sensors. Moreover, heatmaps, with an opacity level that can be set by the user, are loaded and superimposed on the ground orthomap at runtime to show real-time distribution of pollutant, temperature, humidity, etc. Traffic densities are shown as animated 3D crests or arrows to be better visualized in the 3D map. The road graph is queried from the knowledge base and shown over the ground terrain in order to access street information by hovering over the displayed line segments. Both the 3D animated arrows and the road graph can be enabled by the user by selecting the dedicated checkbox in the setting panel of the in-map menu.

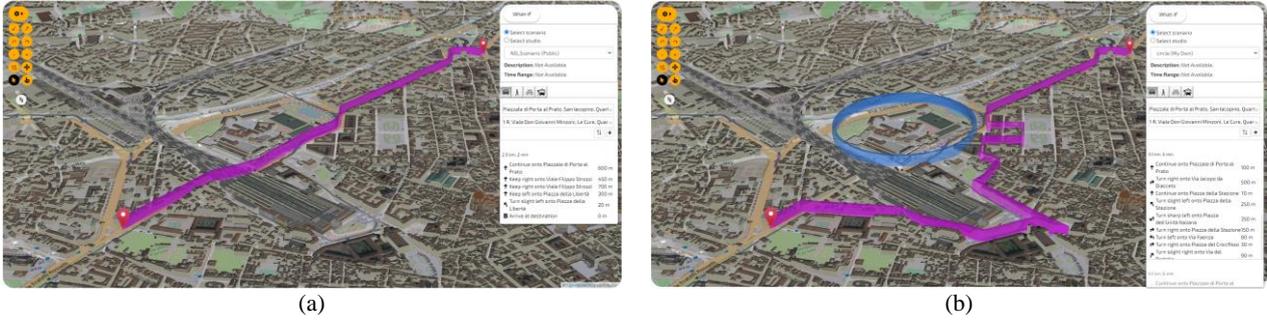

(a) (b)

Figure 11. Example of What-If analysis on vehicle routing. In (a) the initial routing, shown as a purple elevated line. In (b) a What-If scenario is enabled: a blue shape highlights the blocked area, and the updated routing is shown to the user.

In order to show the capability of our SCDT system to let the user carry out simulation and analysis on routing, controls for run What-If analysis were implemented into the web interface. By using a separate interface[*], the user can select specific points or areas to simulate a traffic restriction. Then the scenario can be loaded in the SCDT interface (activating the What-If selector), and the routing algorithm produces trajectories between any start and end position considering the defined restriction. An example of What-If on routing is presented in **Figure 11**: as can be seen, an area was selected to ban traffic from the enclosed streets and the updated routing is shown to the user.

To summarize, our SCDT of Florence includes 3D structures of the urban environment together with static and real-time data describing the status of the city. Simulations are possible by exchanging buildings and performing What-If analysis. Thanks to the developed 3D web engine, the SCDT can be distributed though web browser, requiring neither additional plugins, nor high-end hardware, therefore being accessible from a wide audience enabling possible co-working in virtual living lab and helping to include also common citizen in the urban evolution process.

## 7. Conclusions

In this paper, the Snap4City Smart City Digital Twin framework was presented. According to a series of requirements on field interoperability, data and computing for representations, and distribution and interaction, the proposed framework was designed and developed to aggregate different kinds of data and seamlessly represent them on a freely accessible interactive web interface, therefore realizing an accessible, integrated, replicable, and affordable SCDT solution. The provided solution has been designed taking into account a large range of requirements on: Field Interoperability, Data and computing for representation, and Distribution and interaction. The requirements resulted to be much wider and precise with respect to those of [24] and [27]. The paper also reported an in deep comparative analysis with respect to a large number of smart city digital twin solution in the literature and implemented. Differently from other state-of-the-art SCDTs, the proposed framework exploits the capabilities of the Snap4City IoT platform to ingest, manage, index, and provide through APIs data describing the urban environment that can be continuously updated with real-time information. Our solution exploits exchangeable 3D building models and additional 3D entities together with an accurate terrain model textured with user selectable orthomaps and heatmaps, and information on specific paths and areas (e.g., road graph, cycling paths, etc.) to represent the urban infrastructures. Different entities like IoT devices, POIs, and services are represented as dynamic PINs that can be clicked to access to additional static, historic and real-time data. Specifically devised 3D representations (i.e., 3D crest, arrows, and columns) are used to show real-time traffic density reconstruction or specific sensors. Moreover, the system support What-If analysis tools to modify the actual context producing possible scenarios and simulate the impact of the introduced changes. The paper presented the novel high performance data distribution and rendering engine of the SCDT which permits the access to complex 3D SCDT representations from regular browsers. The new modelling and distribution engine is based on the concept of FusionLayer that offers better performance reducing server requests and lowering client resource usage. It substituted the former data layered structure of the interactive web interface. All the data were integrated into a web interface, using novel client-side business logics based on a layered structure to load all the elements independently with their own safe context avoiding reciprocal interferences and exploiting a tiled subdivision supporting smart caching to deal with huge volumes of data, reduce server calls, and improve the final user experience. The developed framework was used to implement the SCDT of Florence, Italy, to demonstrate all the devised functionalities, and realizing the only freely available SCTD able to represent a wide range of data, perform simulations, and provide the user a large set of tools to inspect, assess, and study the urban environment.

---

[*] Visit
https://www.snap4city.org/dashboardSmartCity/view/index.php?iddasboard=MjE5MA== and activate the Scenarios selector to define a new scenario for What-If analysis.

**Author contributions**

L. Adreani: Methodology, Software, Validation, Writing – Review & Editing.
P. Bellini: Data Curation.
M. Fanfani: Conceptualization, Methodology, Software, Validation, Writing – Original Draft, Writing – Review & Editing, Project administration.
P. Nesi: Conceptualization, Methodology, Validation, Writing – Review & Editing, Supervision, Funding acquisition, Project administration.
G. Pantaleo: Software, Validation.

**Acknowledgements**

The authors would like to thank the MIUR, the University of Florence and the companies involved for co-founding the National Center on Sustainable Mobility, MOST (https://www.centronazionalemost.it/ ). A thanks to the many developers working on the Snap4City platforms. Snap4City (https://www.snap4city.org ) and Km4City are open technologies of DISIT Lab.

Scalable Computing and Communication, IEEE SCALCOM 2019, Leicester, UK
https://www.slideshare.net/paolonesi/data-flow-management-and-visual-analytic-for-big-data-smart-cityiot